\documentclass[aps,prl,twocolumn,showpacs,superscriptaddress]{revtex4-1}
\usepackage{graphicx}
\usepackage{amsmath}
\usepackage{textcomp}
\begin{document}

\title{Tighter spots of light with superposed orbital angular momentum beams}
\author{Pawe{\l} Wo{\'z}niak}
\affiliation{Max Planck Institute for the Science of Light, D-91058 Erlangen, Germany}
\affiliation{Institute of Optics, Information and Photonics, Friedrich-Alexander-University Erlangen-Nuremberg, D-91058 Erlangen, Germany}
\affiliation{Department of Physics and Max Planck - University of Ottawa Centre for Extreme and Quantum Photonics, University of Ottawa, Ottawa ON K1N 6N5, Canada}
\author{Peter Banzer}
\email{peter.banzer@mpl.mpg.de}
\affiliation{Max Planck Institute for the Science of Light, D-91058 Erlangen, Germany}
\affiliation{Institute of Optics, Information and Photonics, Friedrich-Alexander-University Erlangen-Nuremberg, D-91058 Erlangen, Germany}
\affiliation{Department of Physics and Max Planck - University of Ottawa Centre for Extreme and Quantum Photonics, University of Ottawa, Ottawa ON K1N 6N5, Canada}
\author{Fr{\'e}d{\'e}ric Bouchard}
\affiliation{Department of Physics and Max Planck - University of Ottawa Centre for Extreme and Quantum Photonics, University of Ottawa, Ottawa ON K1N 6N5, Canada}
\author{Ebrahim Karimi}
\affiliation{Department of Physics and Max Planck - University of Ottawa Centre for Extreme and Quantum Photonics, University of Ottawa, Ottawa ON K1N 6N5, Canada}
\affiliation{Department of Physics, Institute for Advanced Studies in Basic Sciences, 45137-66731 Zanjan, Iran}
\author{Gerd Leuchs}
\affiliation{Max Planck Institute for the Science of Light, D-91058 Erlangen, Germany}
\affiliation{Institute of Optics, Information and Photonics, Friedrich-Alexander-University Erlangen-Nuremberg, D-91058 Erlangen, Germany}
\affiliation{Department of Physics and Max Planck - University of Ottawa Centre for Extreme and Quantum Photonics, University of Ottawa, Ottawa ON K1N 6N5, Canada}
\author{Robert W. Boyd}
\affiliation{Department of Physics and Max Planck - University of Ottawa Centre for Extreme and Quantum Photonics, University of Ottawa, Ottawa ON K1N 6N5, Canada}
\affiliation{Institute of Optics, University of Rochester, Rochester, New York 14627, USA}
\date{\today}

\begin{abstract}
The possibility of focusing light to an ever tighter spot has important implications for many applications and fields of optics research, such as nano-optics and plasmonics, laser-scanning microscopy, optical data storage and many more. The size of lateral features of the field at the focus depends on several parameters, including the numerical aperture of the focusing system, but also the wavelength and polarization, phase and intensity distribution of the input beam. Here, we study the smallest achievable focal feature sizes of coherent superpositions of two co-propagating beams carrying opposite orbital angular momentum. We investigate the feature sizes for this class of beams not only in the scalar limit, but also use a fully vectorial treatment to discuss the case of tight focusing. Both our numerical simulations and our experimental results confirm that lateral feature sizes considerably smaller than those of a tightly focused Gaussian light beam can be observed. These findings may pave the way for improving the resolution of imaging systems or may find applications in nano-optics experiments.
\end{abstract}
\maketitle
\textit{Introduction} \textemdash The utilization of spatially structured light beams has proven beneficial in many fields of optics research (see \cite{Allen1992,Zhan2009,Padgett2011} and references therein). For instance, phase-structured or polarization-tailored light beams, such as scalar Laguerre-Gaussian (LG) beams carrying orbital angular momentum (OAM) or cylindrical vector beams, hold great potential in nanoplasmonics and nanophotonics \cite{DiFrancia1952,Quabis2000,Youngworth2000,Dorn2003,Dorn2003_2,Kindler2007,Zuechner2008,Mojarad2009,Banzer2010,Sancho2012,ZambranaPuyalto2013,Neugebauer2014,Bauer2015,Wozniak2015,Novotny2006}, optical manipulation and trapping \cite{Allen1992,He1995,Meier2006,Nieminen2008,Padgett2011}, optical communication and sensing \cite{Paterson2005,Tyler2009,Malik2012,BergJohansen2015} and many more. Very prominent examples of research areas where structured light has paved the way for unprecedented enhancements are the fields of imaging and nanoscopy \cite{Hell1994,Huang2009,Schermelleh2010}. In the limit of tight focusing of light beams, the electromagnetic field distribution can become highly complex in the focal plane \cite{Richards1959,Quabis2000,Youngworth2000,Dorn2003,Dorn2003_2}. In this context, the occurrence of longitudinally oscillating field components or, more generally, of three-dimensional field distributions gives rise to a variety of interesting effects and phenomena, including spin-to-orbit coupling \cite{Zhao2007}, transverse angular momentum \cite{Aiello2009,Banzer2013,Aiello2015,Bliokh2015}, the creation of complex polarization topologies \cite{Bauer20152} at the nanoscale, and also the possibility of focusing light more tightly, as observed for instance for radially polarized light \cite{DiFrancia1952,Dorn2003}.\\
\indent In this letter, we study both theoretically and experimentally the smallest achievable lateral feature sizes -- in terms of the field structure -- of specially structured light beams created by the superposition of collinearly propagating LG beams of light forming so-called petal beams. In particular, we investigate such beams not only in the limit of paraxial propagation, but also use a fully vectorial treatment to discuss the case of tight focusing. We show that the smallest achievable feature sizes in such beams can reach sub-wavelength dimensions, and they depend on the OAM carried by the superposed light beams.\\
\indent Starting with the simple case of a plane wave impinging on a lens with circular aperture, the intensity pattern formed in the focal plane of the lens can be calculated using scalar diffraction theory as long as the beam is propagating paraxially. It has the shape of an Airy disc described by a first-order Bessel function (see for instance \cite{Novotny2006}). The size of this focal intensity pattern can be quantified by its radius $d_{R}$ measured from the maximum intensity on-axis to the first null of the intensity distribution. $d_{R}$ depends on the numerical aperture (NA) of the lens used for focusing and the wavelength $\lambda$ of a plane wave as follows:
\begin{equation}
d_{R}\approx0.61\frac{\lambda}{\text{NA}}
\text{.}
\label{eq_rayleigh}
\end{equation}
Consequently, this equation can be used to estimate the spot size (in the scalar limit) for a plane wave focused by a lens with circular aperture. We note that Eq.~(\ref{eq_rayleigh}) is also the result of applying the famous resolution criterion of Lord Rayleigh to the imaging of two closely spaced point-like emitters \cite{Abbe1873,Rayleigh1896,Novotny2006}, in which case $d_{R}$ refers to the minimum resolvable distance of the two emitters.\\
\indent As is customary and for simplicity, we will use Eq.~(\ref{eq_rayleigh}) as gauge for feature sizes also in the regime of high NA focusing where the scalar theory has limited validity. To get a more accurate description of this scheme, vectorial diffraction theory must be applied, taking into account the polarization of the illumination and the effect of depolarization upon focusing \cite{Richards1959,Dorn2003_2}.

\begin{figure}
\begin{center}
\includegraphics[scale=0.40]{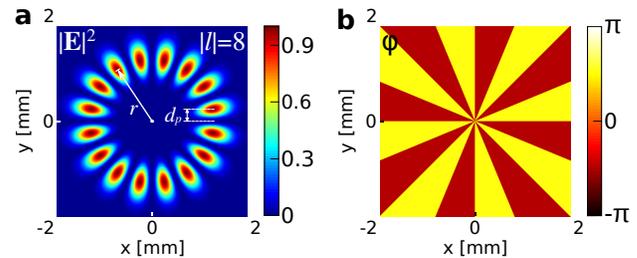}
\caption{\footnotesize (a) Numerically calculated electric energy density (normalized to the maximum value of $\left|\textbf{E}\right|^2$) and (b) phase distributions of an $x$-polarized petal beam for $|l|=8$, $\lambda=535$ nm and $w_0=0.65$ mm.}
\label{fig_petal_beam}
\end{center}
\end{figure}

\begin{figure}
\begin{center}
\includegraphics[scale=0.40]{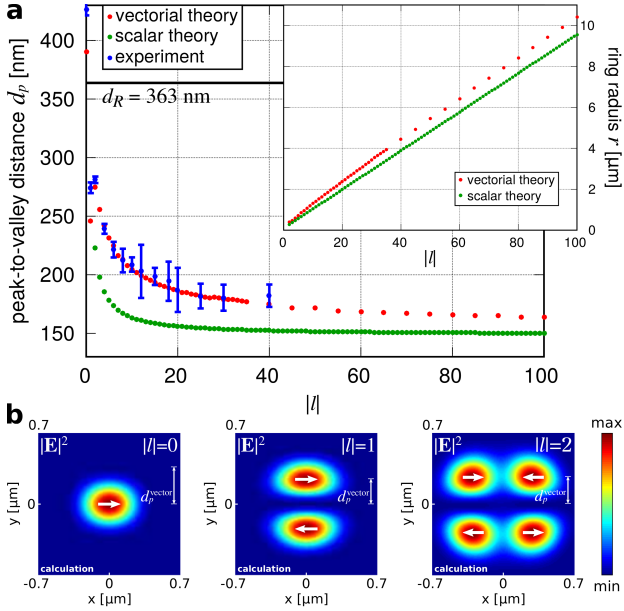}
\caption{\footnotesize (a) Dependence of peak-to-valley distances on $\left|l\right|$ retrieved from the analytical paraxial treatment ($d_{p}^{\text{scalar}}$; green) and from vectorial diffraction theory ($d_{p}^{\text{vector}}$; red) for $\lambda=535$ nm focused with a microscope objective of $\text{NA}=0.9$. The experimental data (blue) is plotted for the experimentally accessible range of values for $\left|l\right|=0\text{--}40$. The inset shows the dependence of the beam radius in the focus on $\left|l\right|$. (b) Distributions of the total electric energy density $\left|\textbf{E}\right|^2$ of tightly focused petal beams for the special cases of $\left|l\right|=0,1,2$.}
\label{fig_petal_distance}
\end{center}
\end{figure}

\textit{Petal beams and their paraxial propagation} \textemdash A multitude of different practical methods capable of generating high-quality light beams carrying orbital angular momentum \cite{Allen2003} have been discussed and demonstrated to date. Amongst others, methods based on cylindrical lenses \cite{Allen1992}, helical phase-plates \cite{Beijersbergen1994} as well as dielectric, liquid-crystal-based or plasmonic q-plates \cite{Marrucci2006,Karimi2014} have been shown. However, the most flexible and tunable generation of LG beams or their superpositions can be realized using spatial light modulators \cite{Heckenberg1992}.\\
\indent If two linearly ($x$-)polarized LG beams of lowest radial order carrying orbital angular momentum of opposite sign ($\text{LG}_{0l}$, $\text{LG}_{0-l}$ for $l\neq0$) co-propagate collinearly, their equally weighted superposition carries no net angular momentum and forms a ring-shaped light beam, consisting of $2|l|$ intensity lobes or petals along the azimuthal coordinate. This is a direct consequence of the inherent azimuthal phase structure of the two beams, resulting in multiple zero-crossings of the electric field along the ring following a $\cos^2(\left|l\right|\phi)$ intensity modulation with $\phi$ the azimuthal coordinate (see Fig.~\ref{fig_petal_beam} for the case of $\left|l\right| = 8$). For the sake of convenience, we have set the radial index of the LG-modes to $0$. Some properties as well as the generation and application of these so-called \textit{cogwheel} or \textit{petal beams} have been studied and discussed in the literature (see for instance \cite{MacDonald2002,Jesacher2004,FrankeArnold2007,Schmitz2008,Naidoo2011,Litvin2013,Krenn2015} and references therein).\\
\indent The half-distance between neighboring petals (peak-to valley distance), equivalent to Eq.~(\ref{eq_rayleigh}), here denoted as $d_{p}$ (see also Fig.~\ref{fig_petal_beam}a), can be easily calculated from the ring radius $r$ and the number $2|l|$ of petals
\begin{equation}
d_{p}=\frac{1}{2}\frac{2\pi r}{2|l|}
\text{,}
\label{eq_petal}
\end{equation} 
for $\left|l\right| \geq 2$. Strictly speaking, this geometrical derivation of the half-distance between neighboring petals is only correct in the limit of large values of $\left|l\right|$ or equivalently a large number of petals formed along the ring-like petal beams. For small values of $\left|l\right|$, $d_{p}$ overestimates the actual half-distance because it measures it along an arc and not along a straight line. Therefore, we restrict Eq.~(\ref{eq_petal}) to cases where $\left|l\right| \geq 2$, because with the case $\left|l\right| = 2$, the beams start to form ring-like distributions of petals.\\
\indent At first glance one might be tempted to believe that by reducing the radius $r$, e.g., by focusing the petal beam, or alternatively by keeping the beam radius fixed but increasing the number of petals $2\left|l\right|$, arbitrarily small values of $d_{p}$ could be achieved. However, an arbitrarily large improvement is not possible as we will show below. From \cite{Saghafi1998,Alonso2011,Padgett2015} it is known that for a paraxial beam $\text{LG}_{0l}$ and, hence, also for the superposition of LG beams as discussed here, the divergence angle $\alpha$ is a function of the beam radius $r$, the wavelength $\lambda$, and the absolute value of the phase charge $\left|l\right|$ carried by the individual beams in the superposition, and reads
\begin{equation}
\tan\alpha=(\left|l\right|+1)\frac{\lambda}{2\pi r}
\text{.}
\label{eq_angle}
\end{equation}
It is worth noting here that the divergence angle $\alpha$ of the beam under investigation might become already considerably large for high values of $\left|l\right|$, even for beam radii $r \gg \lambda$, hence necessitating a non-paraxial treatment. However, it is still very instructive and convenient to retrieve an analytical expression for $d_{p}$ for the paraxial regime. In the paraxial case, the small-angle approximation $\tan\alpha \approx \alpha \approx \sin\alpha = \text{NA}$ can be applied. Together with Eqs.~(\ref{eq_rayleigh}) and (\ref{eq_angle}), Eq.~(\ref{eq_petal}) can be rewritten for the paraxial case and $\left|l\right| \geq 2$ to now read
\begin{equation}
d_{p}^{\text{scalar}} \approx \frac{\left|l\right|+1}{4\left|l\right|}\frac{\lambda}{\text{NA}} = \frac{\left|l\right|+1}{2.44\left|l\right|} d_{R}
\text{.}
\label{eq_r_petal}
\end{equation} 
For the limiting case $\lim_{\left|l\right|\rightarrow\infty}d_{p}^{\text{scalar}}=\frac{d_{R}}{2.44}$. Consequently, the smallest feature sizes of a paraxially propagating petal beam given by the peak-to-valley distance (half the petal spacing) can be smaller than the limit $d_{R}$ as defined in Eq.~(\ref{eq_rayleigh}) by a factor of $\frac{1}{2.44}$. This first simple result shows that by taking advantage of the azimuthal phase structure of LG beams and the interference of two such beams carrying phase vortices of opposite sign, the smallest observable feature sizes along the azimuthal direction can indeed be reduced considerably. Nonetheless, this simple and straight-forward calculation immediately demonstrates that $d_{p}^{\text{scalar}}$ cannot reach arbitrarily small values, even not in the framework of scalar treatment. The value of $d_{p}^{\text{scalar}}$ saturates very quickly for increasing values of $\left|l\right|$ (see also Fig.~\ref{fig_petal_distance}a for the case of high NA). In addition, it should be noted here that this reduction of petal sizes along the azimuthal direction also causes an increase of the radial petal size (see Fig.~\ref{fig_petal_beam}).

\begin{figure}
\begin{center}
\includegraphics[scale=0.40]{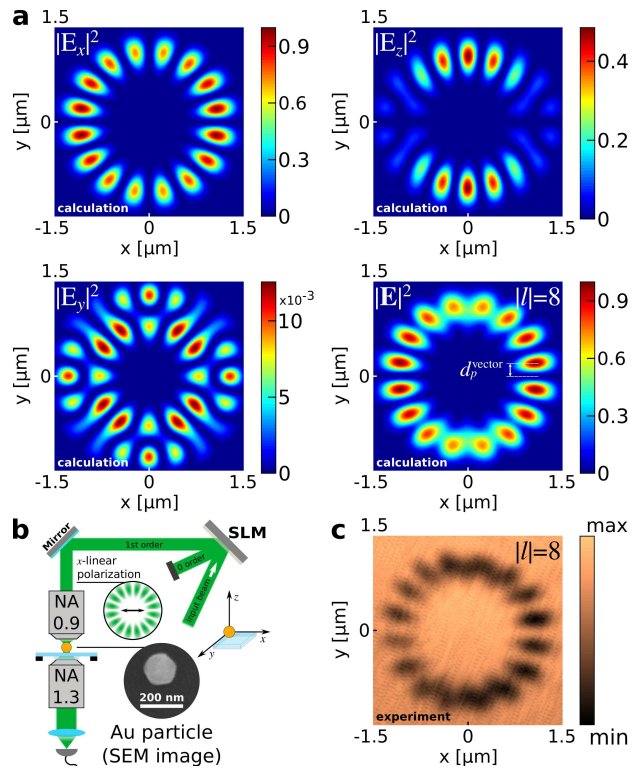}
\caption{\footnotesize (a) Focal distributions of the electric energy density of the individual electric field components $|\text{E}_x|^2$, $|\text{E}_y|^2$ and $|\text{E}_z|^2$ as well as the total electric energy density $|\textbf{E}|^2$ of a tightly focused petal-like beam with $\left|l\right|=8$, $\lambda=535$ nm, $w_0=0.65$ mm (see also Fig.~\ref{fig_petal_beam}). All distributions were numerically calculated using vectorial diffraction theory \citep{Richards1959} and normalized to the maximum of the total electric energy density. (b) Schematic of the experimental setup (similar to Ref. \citep{Banzer2010}). An $x$-polarized Gaussian beam of wavelength $535$ nm is converted into a petal beam by means of a liquid-crystal-based spatial light modulator (SLM; phase-only). The beam is focused by a microscope objective with a numerical aperture (NA) of $0.9$. A $150$--nm diameter gold bead is raster scanned using a three-dimensional piezo-stage through the focal plane of the light beam to determine the focal distribution. The transmitted and forward-scattered light is collected by a high-NA (1.3) oil-immersion lens and measured with a photodiode. (c) An experimental scan showing the total electric energy density distribution of the beam under study as shown in (a).}
\label{fig_focal_fields}
\end{center}
\end{figure}

\textit{Tight focusing of petal beams} \textemdash As already mentioned above, the divergence angle of petal beams, or equivalently $\text{LG}_{0l}$ beams in the superposition, depends on the value $\left|l\right|$ \citep{Saghafi1998,Alonso2011,Padgett2015}. Therefore, it seems to be appropriate to also study the parameter $d_p$ in the limit of strong focusing, using a fully vectorial and nonparaxial description \cite{Richards1959}. In this context it is well-known that upon tight focusing of a light beam, its spatial field distributions may change significantly \citep{Richards1959,DiFrancia1952,Dorn2003,Bauer20152}, accompanied by the appearance of longitudinal field components, i.e. components of the electric field oscillating along the propagation direction. This shows that the polarization state and spatial structure of a light beam has a crucial influence on the shape and size of the focal spot.\\
\indent Using vectorial diffraction theory, we therefore calculated the focal field distributions of tightly focused linearly $x$-polarized petal beams. Without the loss of generality, we only show the components of the electric field here. Fig.~\ref{fig_focal_fields}a shows the corresponding numerical results of the focal field distributions for the input beam shown in Fig.~\ref{fig_petal_beam}, focused with a microscope objective of $\text{NA}=0.9$. The relative size of the input beam for this example ($\left|l\right| = 8$) and for all other cases of $\left|l\right|$ was chosen such that the corresponding beams filled the entrance aperture of the objective lens (for $\left|l\right| = 8$, beam waist $w_0=0.65$ mm). The total electric energy density distribution $\left|\textbf{E}\right|^2$ still shares some similarities with the original input beam, but the intensity lobes along the ring are not equally strong anymore. In addition and even more importantly, the visibility of the intensity lobes located close to the $y$-axis is significantly reduced. Both effects are dominantly caused by the appearance of longitudinal field components peaking on the $y$-axis, where the dominant $x$-component of the electric field is zero (see corresponding distribution of $\left|\text{E}_{z}\right|^2$ in Fig.~\ref{fig_focal_fields}a). In addition, a comparatively weak crossed in-plane component $\left|\text{E}_{y}\right|^2$ is also observed. All distributions are shown in Fig.~\ref{fig_focal_fields}a. It is also worth noting here that, if the input beam was chosen to be circularly polarized, the longitudinal electric field component would form a symmetric ring of intensity lobes, consequently reducing the visibility of all lobes in the total electric energy density along the full ring \cite{Krenn2015}. By choosing linearly polarized light at the input, this problem can be circumvented and regions of optimum visibility with respect to the electric field distribution can be found as discussed above.\\
From the numerically calculated focal field distributions, we now retrieve the peak-to-valley distances in the regions of highest visibility (distance measured from the $x$-axis to the neighboring electric field maximum of $|\textbf{E}|^2$), here denoted as $d_{p}^{\text{vector}}$ (see also Fig.~\ref{fig_focal_fields}a). Alternatively, the circumference of the ring-like intensity distributions can be retrieved from the calculated data and, subsequently, divided by twice the number of petals, as it was done in the paraxial treatment discussed above.\\
\indent Because the total electric energy density or the electric field intensity distribution for small $\left|l\right|$-values evolves from a single or dual-lobe pattern into a ring-like distribution of intensity maxima, we need to define how the values of $d_{p}^{\text{vector}}$ are determined for those cases. In Fig.~\ref{fig_petal_distance}b, we therefore show the corresponding focal distributions of $|\textbf{E}|^2$ for the cases of $\left|l\right|$ = 0 (fundamental Gaussian beam; $x$-polarized $\text{HG}_{00}$ beam), $\left|l\right| = 1$ (first order Hermite-Gaussian beam; $x$-polarized $\text{HG}_{01}$ beam) and $\left|l\right| = 2$ ($x$-polarized $\text{HG}_{11}$ beam). For a linearly polarized Gaussian beam ($\left|l\right|$ = 0), the distance $d_{p}^{\text{vector}}$ is measured from the optical axis to the first zero-crossing of the electric field along the $y$-axis perpendicular to the input polarization (see Fig.~\ref{fig_petal_distance}b). Similarly, we retrieve this value for the cases $\left|l\right| = 1$ and $\left|l\right| = 2$.\\
\indent The retrieved values of $d_{p}^{\text{vector}}$ as a function of $\left|l\right|$ are plotted in Fig.~\ref{fig_petal_distance}a (red) with a step size of $1$ ($5$) for $\left|l\right| < 35$ ($\left|l\right| \geq 35$) and $\text{NA}=0.9$, $\lambda=535$ nm, together with $d_R$ calculated from Eq.~(\ref{eq_rayleigh}) (black solid line). The beam waist parameter $w_0$ for each input beam was chosen such that the outer radius, at which the modulus of the electric field $\left|\textbf{E}\right|$ reaches a value of $\text{max}(|\textbf{E}|)/e$, was coinciding with the edge of the entrance aperture of the focusing lens (see sketch in Fig.~\ref{fig_focal_fields}b). In addition, we also include the corresponding values of $d_{p}^{\text{scalar}}$ (see also Eq.~\ref{eq_r_petal}) for $\left|l\right| \geq 2$, resulting from the scalar treatment (green), here applied to the case of high NA.\\
As can be seen, the values of $d_{p}^{\text{scalar}}$, which result from the scalar, fully paraxial and analytical treatment of the petal beams, predict smaller peak-to-valley distances in comparison to the actual distances $d_{p}^{\text{vector}}$ retrieved from a fully vectorial calculation. In fact, the same holds true for $d_{p}^{\text{vector}}|_{\left|l\right| = 0}$ (fundamental Gaussian beam), which can be compared to $d_R$. Also here, the approximate scalar theory (see Eq.~(\ref{eq_rayleigh})) predicts a smaller value. However, in both theoretical cases, the peak-to-valley distances decrease with increasing ${\left|l\right|}$ before they eventually saturate. This analysis reveals that for the wavelength and the focusing parameters used in this example, the peak-to-valley distance reaches a minimum value $d_{p}^{\text{vector}} \approx \frac{1}{2.25} d_{R}$ for ${\left|l\right|} \gg 1$. This minimum value was retrieved from an extrapolation of the data for $d_{p}^{\text{vector}}$ shown in Fig.~\ref{fig_petal_distance}a. As mentioned already, this limit is different to the aforementioned value $d_{p}^{\text{scalar}} = \frac{1}{2.44} d_{R}$ for large values of ${\left|l\right|}$, resulting from the paraxial treatment. In Fig.~\ref{fig_petal_distance}a (inset), we also plot the dependence of the beam (ring) radius $r$ on ${\left|l\right|}$. For both theoretical treatments (paraxial and nonparaxial), a linear dependence of the beam radius on ${\left|l\right|}$ is found. In other words, the beam diameter or radius is growing linearly with increasing ${\left|l\right|}$. However, the smallest observable feature sizes of the beam, $d_{p}^{\text{vector}}$ and $d_{p}^{\text{scalar}}$ (petal sizes), in both regimes decrease with increasing ${\left|l\right|}$ until they reach a limit, which is far below the value of $d_{R}$ in Eq.~(\ref{eq_rayleigh}). From this perspective, Eq.~(\ref{eq_r_petal}) can be seen as an adapted version of Eq.~(\ref{eq_rayleigh}) for petal beams taking into account the spatial structure of light. It is worth noting here that the observed minimal lateral feature size for petal beams of the shown type are even considerably smaller than for those cases discussed in the literature, for instance for tightly focused radially polarized light beams \cite{Dorn2003}, and they show higher visibility.

\textit{Experimental realization} \textemdash To experimentally verify our theoretical findings, we also performed scan measurements for tightly focused petal beams ($\left|l\right|=\left[0,40\right]$). For that purpose, a custom-built scanning setup was used \cite{Banzer2010} (Fig.~\ref{fig_focal_fields}b). A linearly $x$-polarized Gaussian beam at a wavelength of $\lambda=535$ nm was converted into a petal beam, i.e. a superposition of two collinearly propagating $\text{LG}_{0l}$ and $\text{LG}_{0-l}$ beams, using a liquid-crystal-based phase-only reflective spatial-light-modulator (SLM) \cite{Heckenberg1992,DAmbrosio2013,Bolduc2013} in a single-pass configuration. The beam was then focused with a microscope objective ($\text{NA}=0.9$). The focal electric energy density distribution was raster-scanned by a gold nanoparticle of diameter $150$ nm placed on a glass substrate. To be able to compare the experimental results with the numerical calculations based on the vectorial diffraction theory, the waist of the input beam was chosen appropriately to fill the aperture of the objective lens, as described before. The focused beam was scanned by changing the position of the gold nanoprobe using a three-dimensional piezo-stage. For each position of the particle relative to the optical axis within the focal plane, the transmitted and forward-scattered light was collected by a second oil-immersion objective lens ($\text{NA}=1.3$) and measured with a photo-diode. Both microscope lenses were aligned confocally. The chosen plasmonic sub-wavelength particle is perfect for probing the local electric field. While scanning the beam, the particle is excited depending on its relative position. Consequently, a raster-scan measurement results in a two-dimensional scan image, where a signal decrease is in first approximation proportional to the local electric energy density and, therefore, containing information about its spatial focal distribution. It should be noted here that for the chosen wavelength and particle size, a significant contribution of a quadrupole to the scattering of the particle is expected. Furthermore, a more advanced experimental method as described in Ref.~\citep{Bauer2014} should be used if the full field information (amplitudes and phases of individual field components) of the beams under study need to be measured. In Fig.~\ref{fig_focal_fields}c, we show the experimental result of a petal beam for $\left|l\right|=8$ (equivalent to the case shown in Fig.~\ref{fig_focal_fields}a). The experimental scan result is in very good agreement with the numerically calculated distribution shown in Fig.~\ref{fig_focal_fields}a. The fact that the longitudinal field components appear more pronounced in comparison to the transverse field components is a direct consequence of the different collection efficiencies (defined by the lower objective) for light emitted from a longitudinally or transversally oscillating dipolar or quadrupolar modes excited in the particle.\\
\indent Following the aforementioned procedure, measurements for beams and petal beams with $\left|l\right|=\left[0,40\right]$ were performed, and the peak-to-valley distances were retrieved in the regions of highest visibility, equivalent to the retrieval of $d_{p}^{\text{vector}}$. The corresponding data is plotted in Fig.~\ref{fig_petal_distance}a (blue). The experimentally retrieved values are also in very good agreement with the data obtained before, using vectorial diffraction theory.

\textit{Conclusions} \textemdash In our study, we examine the spatial distribution of the electric energy density of tightly focused petal beams both theoretically and experimentally. Both the scalar as well as the fully vectorial theoretical treatment of the investigated scheme predict a significant reduction of the minimum observable feature sizes (peak-to-valley distances) in comparison to the case of a plane wave. This theoretical part of our study emphasizes the importance of a fully vectorial and nonparaxial theory for describing the propagation of the beams under investigation, especially for large values of $\left|l\right|$. In addition, our experimental study, based on a nanoparticle utilized as a scanning probe, confirms the theoretical predictions. Such beams might find interesting applications in nano-optics and plasmonics. If the field distribution of the input beams superposed to form a petal beam was also tailored in polarization in addition to the phase, the focal field could be modified even further.

\textit{Acknowledgements} \textemdash The authors thank A. Zeilinger and M. Krenn for interesting discussions. PB acknowledges financial support by the Alexander von Humboldt Foundation and the Canada Excellence Research Chair (CERC) in Quantum Nonlinear Optics.
\bibliography{biblioteca}
\end{document}